\documentclass{emulateapj} 
 
\shorttitle{High energy emission} \shortauthors{Peer \& Waxman} 
  
\newcommand{\eV}{\rm{\, eV}} 
\newcommand{\keV}{\rm{\, keV}} 
\newcommand{\MeV}{\rm{\, MeV}} 
\newcommand{\GeV}{\rm{\, GeV}} 
\newcommand{\TeV}{\rm{\, TeV }} 
 
\newcommand{\flu}{\rm{\, erg \, cm^{-2} \, sec^{-1}}} 
\newcommand{\beq}{\begin{equation}} 
\newcommand{\eeq}{\end{equation}} 
\newcommand{\ba}{\begin{array}} 
\newcommand{\ea}{\end{array}} 
 
\begin{document} 
 
\title{High energy photon emission in the early afterglow of GRB's} 
 
\author{Asaf Pe'er\altaffilmark{1}\altaffilmark{2} and Eli Waxman\altaffilmark{1}} 
\altaffiltext{1}{Physics faculty, Weizmann Institute of Science, 
Rehovot 76100, Israel} 
\altaffiltext{2}{asaf@wicc.weizmann.ac.il}

\begin{abstract} 
 
We consider the emission within the fireball model framework of very
high energy, $\sim1$~GeV to $>1$~TeV photons, on a minute time scale,
during the onset of fireball deceleration due to interaction with
surrounding medium. Our time dependent numerical model includes    
exact treatment of electron synchrotron emission, inverse-Compton
scattering, pair production, and  
evolution of electromagnetic cascades (initiated by pair production or
photo-production of pions). We find that (i) The 1~GeV--10~GeV flux is
not sensitive to model parameters and is $\sim 10^{-7} \flu$ for $z=1$
bursts, well within the detection capabilities of GLAST; (ii) The
sub-TeV flux depends on the surrounding medium density and on the
fraction of thermal energy carried by the magnetic field,
$\epsilon_B$: It ranges from $\sim 10^{-7} \flu$ in the case of
typical ISM density and $\epsilon_B \lesssim 10^{-4}$ to $10^{-10}
\flu$ in the case of a source surrounded by a wind and
$\epsilon_B\sim10^{-0.5}$;  
(iii) The sub-TeV flux is detectable by high energy $\gamma$-ray 
experiments such as HESS, MAGIC, MILAGRO, and VERITAS; 
(iv) Combined $\sim1 \keV$, $\sim1 \GeV$ and sub-TeV observations will allow 
to determine both $\epsilon_B$ and the ambient medium density; 
(v) The spectra depend only weakly on the spectral index of 
the energy distribution of the accelerated electrons. 
Pion production energy loss of high energy protons may contribute
significantly in the wind case to the luminosity of high energy
photons. However, it is difficult to distinguish in this case between
the electron and proton contributions since the spectral shape is
determined primarily by the energy dependence of the pair production
optical depth. 
\end{abstract}

 
\section{Introduction} 

In the standard fireball model of gamma-ray-bursts (GRB's), the
observable effects are due to the dissipation  of kinetic energy in a
highly relativistic fireball
\citep[for reviews, see, e.g.,][]{fireballs1, fireballs2, W03}. 
The observed radiation is well explained as synchrotron and
inverse-Compton emission from shock accelerated electrons.   
Electrons accelerated in internal shock waves within the expanding
fireball produce the prompt $\gamma$-ray emission, 
while electrons accelerated in the external shock wave driven by the
fireball into the surrounding medium produce the  
afterglow emission, from the X to the radio bands,  
that can last months after the burst \citep{PR93, 
  MR97, Viet97, W97a, W97b, Sari98, GW99}.   

At the early stages of the afterglow, as the energy density of the  
surrounding matter swept by the external shock wave equals the  
energy density of the fireball plasma, 
a reverse shock crosses 
the fireball plasma \citep{MRP94, SP95}.  
At this transition phase, lasting over a duration comparable  
to that of the burst itself, both forward and reverse shocks accelerate particles, 
which subsequently emit the early afterglow.  
Early optical emission from several GRBs has been identified as reverse 
shock emission (\citet{Akerlof99,SP99,MR99,Kob00,KS00,Wei03,ZKM03}).         

In the past few years, evidence accumulated for the existence of a
high energy  emission component ($\sim 1\GeV - 1\TeV$) associated with
gamma-ray bursts (GRBs),  which are typically observed at much lower
energies, few keV to few MeV.  Such a component was first observed by
EGRET in the $\sim 100 \MeV - 18 \GeV$ range
\citep{Schneid92,Hurley94}. There is evidence for even higher energy, 
$\sim1 \TeV$ emission, observed in one burst out of 54 BATSE GRBs in
the MILAGRITO field of view  \citep{Atkins00, Atkins03}.  
 
The mechanism responsible for high energy photon emission is still controversial.  
High energy electrons inverse Compton (IC) scatter low energy photons 
to high energies, 1\GeV - 1\TeV  \citep{MRP94} extending the spectrum
to high energy. 
The exact shape of the spectrum depends on the  values of the uncertain
parameters of the model \citep{CD99, Wang01, ZM01}.  
An additional source of high energy emission is the  possible
acceleration of baryons, first suggested by \citet{W95}, \citet{Viet95} and \citet{MnU95}.  
Accelerated protons can emit synchrotron radiation at this energy band  
\citep{V97, Totani98a, Totani98b, BD98}.   
High energy baryons can also produce energetic pions,  via photo-meson
interactions with the low energy photons, creating an alternative
source of high energy photons and neutrinos \citep{WB97, WB00, BD98}. 
A high energy component may also be created by the decay of energetic
pions, produced in proton-proton(neutron) collisions, if the expanding
plasma collides with a dense ($n \sim 10^{11} \rm{\, cm^{-3}}$) cloud
\citep{Katz94, PIO00}.   
  
In this paper, we analyze the high energy emission component during
the early afterglow (transition phase) on a time scale of tens to
hundreds of seconds following the GRB, within the framework of the
fireball model. We explore the dependence of photon spectrum on
uncertain model parameters, in particular on the energy density of the
magnetic field, the power law index of the accelerated particles and
the density of the medium surrounding the fireball. The ambient medium
density differs widely between different scenarios for GRB
production. In the neutron star merger scenario \citep{Good86,Pac86,
  Eichler89}, for example, a density typical to the ISM ($n \approx 1
\rm{\, cm^{-3}}$) is generally expected, while in the massive stellar
collapse scenario \citep{Woosley93, Levinson93, Pac98} a much higher
density ($n \approx 10^3 - 10^4 \rm{\, cm^{-3}}$) may be expected in a
wind surrounding the progenitor, due to mass loss preceding the
collapse \citep{Chevalier01}.  In addition, we explore the possible
contribution of high energy protons.
 
Calculation of the early high energy emission spectrum is complicated
due to several reasons: First, IC scattering is partly in the
Klein-Nishina regime. Second, high energy photons (produced by IC
scattering or via pion decay) initiate high energy electro-magnetic
cascades, the evolution of which is difficult to study analytically.
 Studies of cascade processes in the past
 ( \citet{BR71,Guilbert83,Svensson87}) have shown that it may have a
  significant effect on the spectrum at high energies.
And third, the optical depth to pair production may be large at high
energy, leading to the formation of a large number of $e^\pm$ pairs
which affect the resulting spectra. Due to these complications,
numerical calculations are required to accurately determine the high
energy spectra and their dependence on uncertain model
parameters. 

Numerical methods where extensively used in the past in the study
of active galactic nucleus (AGN) plasma \citep{Guilbert83, ZL85, FBGPC,
LZ87, Coppi92}. Using these methods, new results were obtained, such
as the complex pattern of the spectral indices in the X-ray ($2-10
\keV$) range, that were not obtained by previous analytic calculations.
In the context of the high energy emission from GRB's, numerical
calculations are challenging. The 
large difference between the characteristic time scales for the
evolution of high energy and low energy particles, and the fact that
the distribution of low energy particles evolves on a time scale
comparable to the dynamical time and does not reach a steady-state,
implies that the numerical code must follow processes characterized by
widely differing time scales. Due to these difficulties, the accuracy
of existing numerical calculations of emission from GRB's \citep{Pan98} is
limited above 
$\sim1 \GeV$. Here we present numerical results using a code that
allows to overcome the difficulties mentioned above. Our time
dependent model includes an accurate description of synchrotron
emission, direct and inverse Compton scattering, pair production and
the evolution of high energy electromagnetic cascades. 

Several new high energy photon detectors will become fully operational
in the near future. These include the
GLAST\footnote{http://www-glast.stanford.edu} satellite, which will
greatly improve the sensitivity at $\sim1 \GeV$ and will open a new
window of observation up to $\sim100$~GeV.
GLAST sensitivity, $\sim 10^{-13} \flu$ at 1~GeV will allow detection
of many hundreds of bursts per year, if GRB's emit equal amount of
energy at GeV and keV bands. 
In addition, new generation of
sub-TeV Cerenkov detectors such as
MAGIC\footnote{http://hegra1.mppmu.mpg.de/MAGICWeb},  
HESS\footnote{http://www.mpi-hd.mpg.de/hfm/HESS/HESS.html}
VERITAS\footnote{http://veritas.sao.arizona.edu/} and CANGAROO
III\footnote{http://icrhp9.icrr.u-tokyo.ac.jp/} will open new era in
GRB observations. The universe is  
transparent to 100~GeV photons up to redshift of $\approx0.5$
\citep[e.g.][]{SS98}. Given the local rate of GRBs \citep{GPW03}, 
the sensitivity of these detectors at $\sim100 \GeV$, $\sim10^{-10}
\flu$ for a 100 second burst, will allow a detection of few to few
tens of GRB's per year at this energy band (depending on the exact
instrument energy threshold) if the GRB spectrum in the keV - sub TeV range is
flat, $\nu F_\nu \propto \nu^0$. The field of view of the Cerenkov
telescopes is small, which implies that fast slewing to the GRB
position, on minute time scale, would be required. This may be
achievable with the fast alerts that will be provided by
SWIFT\footnote{http://www.swift.psu.edu/}. The sensitivity of the
MILAGRO\footnote{http://www.lanl.gov/milagro/} shower detector at
$\sim1$~TeV would allow it to detect $\sim1$ event per year.  The
results presented here may therefore be useful for planning the
observing strategy of the detectors and may allow to use high energy
data to constrain the values of uncertain parameters of the model.  

This paper is organized as follows. In \S\ref{sec:model} we derive the
plasma conditions during the transition phase, which is the duration over
which the reverse shock exists. We calculate the
critical synchrotron frequencies and luminosities in the "wind"
scenario (\S\ref{sec:model_wind}) and in the "uniform density ISM"
scenario (\S\ref{sec:model_nw}). In \S\ref{sec:prot} we discuss proton
energy loss and its contribution to high energy photon emission. 
Our numerical model is briefly presented in
\S\ref{sec:numerics};  A detailed description of the model may be
found in \citet{PW03}. In \S\ref{sec:results} we present our numerical
results. We summarize and conclude in \S\ref{sec:discussion},  with
special emphasis on implications for high energy photon telescopes.

\section{Model assumptions, plasma conditions and proton energy loss  
at the transition phase} 
\label{sec:model} 
 
The interaction of fireball ejecta with surrounding gas produces a
reverse shock which propagates into and decelerates the fireball
ejecta. As the reverse shock crosses the ejecta, it erases the memory
of the initial conditions, and the expansion then approaches the
Blandford-McKee self-similar solutions \citep{BM76}.  We derive in
this section the plasma parameters, the low energy photon luminosity
and spectrum, and the proton energy loss time scales expected during
this transition phase, for fireball expanding into a wind and into a
uniform medium with density typical to the ISM. 
 
\subsection{Wind scenario} 
\label{sec:model_wind}
 
The mass loss preceding a collapse creates a density profile 
$\rho(r)= A r^{-2}$, where the proportionality constant is  
taken as $A=5\times10^{11} A_* \rm{\,g\,cm^{-1}}$,  
and a value of  $A_* \approx 1$ is assumed for a typical  Wolf-Rayet 
star \citep{Willis91}. 
During the self similar expansion, the Lorentz factor of plasma
behind the forward shock is  
$\Gamma_{\rm BM} = \left(9 E / 16 \pi \epsilon(r) \right)^{1/2} r^{-3/2}$ , 
where $\epsilon(r)$ is the  
energy density of the surrounding medium,  
$\epsilon(r) \approx \rho(r) c^2$, 
and $E$ is the (isotropically equivalent) fireball energy. 
The characteristic time at which radiation emitted by shocked plasma  
at radius $r$ is observed by a distant observer is  
$t \approx r/4 \Gamma_{\rm{BM}}^2 c$ \citep{W97}. 
 
The reverse shock is relativistic if the   
energy density of the surrounding matter swept by the  
forward shock equals the energy density of the  
propagating plasma. 
Since both energy densities decrease with the second power of the radius, 
a relativistic shock is formed if 
$E/(4 \pi r^2 \Gamma_i^2 cT) \leq 4 \Gamma_i^2 \epsilon(r)$, 
i.e. if 
\beq  
E \leq 16 \pi r^2 \Gamma_i^4 c T \epsilon(r) =  
 5.5\times 10^{56} A_* T_2 \Gamma_{i,2.5}^4 \rm{\,erg}, 
\label{rev_cond} 
\eeq 
where $\Gamma_i = 10^{2.5} \Gamma_{i,2.5}$ is the original ejecta Lorentz factor, 
and $T = 100 \, T_2 \rm{\, s}$ is the burst duration. 
At higher explosion energy, the reverse shock is not relativistic. 
Both prompt and afterglow observations suggest that $E \lesssim 10^{54} \rm{\,erg}$, 
implying that a relativistic reverse shock should in general be formed.

The duration $T$ (measured in the observer frame)  
of the transition phase,  
during which the reverse shock exists, is comparable 
to the longer of the two time scales set by the initial conditions 
\citep {W03}: 
The (observed) GRB duration $T_{GRB}$ 
and the (observed) time $T_\Gamma$ at which 
the self-similar Lorentz factor $\Gamma_{\rm{BM}}$  
equals the original ejecta Lorentz factor $\Gamma_i$, 
$\Gamma_{\rm{BM}} (T_\Gamma) = \Gamma_i$. 
This implies  
that the transition radius to self similar expansion, $r_s=4\Gamma_{\rm{BM}}^2 c T$, is
\beq 
\ba{lll}
 r_s & = \rm{max} &\left(  
2.2\times10^{16} \, E_{53}^{1/2}T_2^{1/2} A_*^{-1/2} ,\right. \nonumber \\
& & \left.4.4\times10^{14} \, E_{53}\Gamma_{i,2.5}^{-2}   A_*^{-1} 
\right) \rm{\,cm}, 
\label{r_max_w} 
\ea
\eeq 
where $E = 10^{53} E_{53} \rm{\, erg}$. 
During the transition, shocked plasma 
expands with Lorentz factor close to that given by the self-similar 
solution, $\Gamma \simeq \Gamma_{\rm{BM}}(r=r_s)$, i.e., 
\beq 
\Gamma \simeq  \left(  
\frac{9 E}{64 \pi A T c^3} \right)^{1/4}  
= 43 \, E_{53}^{1/4} T_2^{-1/4} A_*^{-1/4}. 
\label{G_w} 
\eeq 
The Lorentz factor of the reverse shock in the frame of the unshocked 
plasma is
$\Gamma_r -1 \approx \Gamma_i/\Gamma$. 
  
We denote by  $\epsilon_e$ and  $\epsilon_B$ the fractions of the  
thermal energy density 
that are carried, respectively,  by electrons and magnetic fields.
The characteristic Lorentz factor of electrons accelerated at the 
forward shock is
$\gamma_{char,f} = \epsilon_e \Gamma_{\rm{BM}} m_p/m_e$. 
Electrons accelerated at the reverse shock are characterized by  
$\gamma_{char,r} = \epsilon_e  (\Gamma_r -1) m_p/m_e \simeq
   \epsilon_e (\Gamma_i/\Gamma_{\rm{BM}}) m_p/m_e$.
The minimum Lorentz factor of a power law accelerated electrons is 
\beq 
\gamma_{\min} = \gamma_{char} \times  \left \{  
\ba{ll} 
\log\left( \frac{\gamma_{\max}}{\gamma_{\min}}\right)^{-1} & p=2 
\nonumber \\ 
\left( \frac{p-2}{p-1} \right) & p \neq 2 , 
\ea 
\right. 
\label{eq:gamma_min}
\eeq 
where $p$ is the power law index of the accelerated electrons energy
  distribution, $n_e(\varepsilon_e) \propto \varepsilon_e^{-p}$.

Synchrotron emission peaks at 
$ \varepsilon_{\gamma, m}^{ob.} = (3/2) \Gamma \hbar \gamma_{\min}^2 
q B /m_e c$. Using 
\beq 
u = \sqrt{\frac{4 \pi A^3 c^3}{ 9 E T^3}} =  
7.0 \times 10^3 \, E_{53}^{-1/2} T_2^{-3/2} A_*^{3/2} \rm{\, erg\, cm^{-3}}
\eeq 
for the energy density behind the forward shock at the transition
radius, the synchrotron emission peak at the reverse and forward
shocks is given by
\beq 
\ba {l} 
\varepsilon_{\gamma, m,r}^{ob.} = 0.5\, (1+z)^{-1} \quad E_{53}^{-1/2} T_2^{-1/2} A_* 
\Gamma_{i,2.5}^2 \epsilon_{e,-1}^2 \epsilon_{B,-2}^{1/2} \eV,  \nonumber \\ 
\varepsilon_{\gamma, m,f}^{ob.} = 20  \, (1+z)^{-1} \quad  E_{53}^{1/2} T_2^{-3/2} 
\epsilon_{e,-1}^2 \epsilon_{B,-2}^{1/2} \eV, 
\ea 
\eeq 
where $p=2$, $\log(\gamma_{\max}/\gamma_{\min}) \simeq 10$  
assumed, $\epsilon_e = 0.1\, \epsilon_{e,-1}$, and $\epsilon_B = 0.01\, \epsilon_{B,-2}$. 
These energies are above $\varepsilon_{\gamma, c}$, the characteristic 
synchrotron energy of electrons for which the synchrotron cooling 
time, $9 m_e^3 c^5 / 4 q^4 B^2 \gamma$, is comparable to the ejecta 
(rest frame) expansion time, $\sim r/\Gamma c$,  
\beq 
\varepsilon_{\gamma, c}^{ob.} = 2.1 \times 10^{-2} \, (1+z)^{-1} \quad
E_{53}^{1/2} T_2^{1/2} A_*^{-2} \epsilon_{B,-2}^{-3/2} \eV.  
\eeq   
$\varepsilon_{\gamma, m,r}^{ob.}$ is  
 comparable to the synchrotron self absorption energy, below which the 
optical depth becomes larger than 1, 
\beq 
\varepsilon_{ssa}^{ob.} = 0.4 (1+z)^{-1} \quad \, T_2^{-1} A_*^{2/3} \epsilon_{e,-1}^{1/3} 
\epsilon_{B,-2}^{1/3} \eV
\label{eq:ssa_wind}
\eeq 
(for $p=2$).
Equating the particle acceleration time, $t_{acc} \simeq E/(cqB)$ and
the synchrotron cooling time, gives the maximum Lorentz factor of the
accelerated electrons (in the plasma rest frame),  
$\gamma_{e,\max} = 1.8 \times 10^7 \,
  E_{53}^{1/8} \, T_2^{3/8} \, A_*^{-3/8} \epsilon_{B,-2}^{-1/4}$.
Synchrotron emission from these electrons peaks at
\beq
\varepsilon_{\max}^{ob.} = 1.0 \times 10^{10}   (1+z)^{-1} \quad \,
E_{53}^{1/4} T_2^{-1/4} A_*^{-1/4} \eV. 
\eeq
 
The reverse shock specific luminosity  
$L_\varepsilon = dL/d\varepsilon_\gamma^{ob.}$ at $\varepsilon_{\gamma,m,r}^{ob.}$ is 
\beq 
L_m =  
2.8 \times 10^{60} \, E_{53}^{3/2} T_2^{-1/2} A_*^{-1} 
\Gamma_{i,2.5}^{-2} \epsilon_{e,-1}^{-1} \epsilon_{B,-2}^{-1/2} \rm{\, s^{-1}}.  
\eeq   
For a power law energy distribution with $p=2$ of the accelerated electrons,
the specific luminosity is $L_\varepsilon\propto\varepsilon_\gamma^{-1}$
above $\varepsilon_{\gamma,m,r}$.

\subsection{Uniform density ISM} 
\label{sec:model_nw} 

During self similar expansion into uniform density medium, 
the Lorentz factor of plasma is given by  
$\Gamma_{\rm{\rm{BM}}} = \left(17E/16\pi n m_p c^2 \right)^{1/2} r^{-3/2}$ 
\citep{BM76}. 
The transition to self similar expansion occurs at a radius 
\beq 
\ba{lll}
r_s & = \rm{max}  & \left(  
6.3 \times 10^{16} \, E_{53}^{1/3} n_0^{-1/3} \Gamma_{i,2.5}^{-2/3} ,\right. \nonumber \\
& & \left. 1.3 \times 10^{17} \, E_{53}^{1/4} n_0^{-1/4} T_2^{1/4} \right) \rm{\,cm},  
\ea
\eeq 
where a typical value $n =1 \, n_0 \rm{\, cm^{-3}}$ of the ambient number density assumed.
Calculations of the characteristic Lorentz factors,  
synchrotron frequencies and characteristic luminosities closely  
follow the steps of the wind scenario calculations,  
and partly appear in \citet{WB00}.
The peak energy of synchrotron emission from electrons accelerated at the reverse shock,
\beq 
\varepsilon_{\gamma, m,r}^{ob.} = 2 \times 10^{-2}\, (1+z)^{-1} \quad 
n_0^{1/2} \Gamma_{i,2.5}^2 \epsilon_{e,-1}^2 \epsilon_{B,-2}^{1/2} \eV,  
\eeq 
is comparable to the synchrotron self absorption energy,
\beq 
\varepsilon_{ssa}^{ob.} = 5 \times 10^{-2} (1+z)^{-1} \quad
E_{53}^{1/3} n_0^{1/3} T_2^{-2/3} \epsilon_{e,-1}^{1/3} 
\epsilon_{B,-2}^{1/3} \eV,
\label{eq:ssa_ism}
\eeq 
and is below 
\beq 
\varepsilon_{\gamma, c}^{ob.} = 10 \, (1+z)^{-1} \quad
E_{53}^{-1/2} n_0^{-1} T_2^{-1/2} \epsilon_{B,-2}^{-3/2} \eV.  
\eeq   
Synchrotron emission from electrons accelerated at the forward shock
peaks at higher energy,
\beq 
\varepsilon_{\gamma, m,f}^{ob.} = 25 \, (1+z)^{-1} \quad
E_{53}^{1/2} T_2^{-3/2} \epsilon_{e,-1}^2 \epsilon_{B,-2}^{1/2} \eV.  
\eeq   
The reverse shock specific luminosity,
$L_\varepsilon = dL/d\varepsilon_\gamma^{ob.}$ at $\varepsilon_{\gamma,c}^{ob.}$ is 
\beq 
L_c =  
1.5 \times 10^{59} \, E_{53}^{3/2} T_2^{-1/2} n_0 
\epsilon_{e,-1} \epsilon_{B,-2}^{3/2} \rm{\, s^{-1}}.  
\eeq   
For a power law energy distribution with $p=2$
 of the accelerated electrons,
 $L_\varepsilon\propto\varepsilon_\gamma^{-1}$
at higher energies, $\varepsilon > \varepsilon_{\gamma,c}$.

\subsection{Proton energy loss and its contribution to high energy photon emission}
\label{sec:prot} 

Protons accelerated to high energy in the reverse shock contribute to
the emission of high energy photons by synchrotron emission and
photo-production of pions, which decay to produce high energy photons
and electrons (and neutrinos). The maximum energy to which protons are
accelerated is determined by equating the proton acceleration time (in
the plasma frame), $t_{acc} \simeq \gamma_pm_pc^2/c q B$, to the
minimum of the dynamical time, the synchrotron energy loss time, and
the pion production energy loss time. For expansion into a wind, the
acceleration time is equal to the dynamical time, $T_{dyn} = 4 \Gamma T$, for
protons with (plasma frame) Lorentz factor 
\beq 
\gamma_{p,dyn} \approx 2 \times 10^{10} \, A_*^{1/2} \epsilon_{B,-1}^{1/2},
\label{gamma_p_M} 
\eeq   
and to the synchrotron loss time, $t_{syn} = 9 m_p^3 c^5 / 4 q^4 B^2 \gamma_p$, for protons with (plasma frame) Lorentz factor
\beq 
\gamma_{p,syn} \approx 2 \times 10^{10} \, E_{53}^{1/8}T_2^{3/8}
A_*^{-3/8} \epsilon_{B,-1}^{-1/4}.
\label{gamma_syn} 
\eeq   

The time for energy loss via pion production may be approximated as \citep{WB97} 
\beq 
t_\pi^{-1}  = \frac{1}{2\gamma_p^2}c \int_{\varepsilon_0}^\infty 
d\varepsilon \sigma_\pi (\varepsilon) \xi (\varepsilon) \varepsilon  
\int_{\varepsilon/2 \gamma_p}^\infty dx x^{-2} n(x).     
\label{eq:t_pi_int}
\eeq 
$\sigma_\pi(\varepsilon)$ is the cross section for pion production 
for a photon with energy $\varepsilon$ in the proton rest frame, 
$\xi(\varepsilon)$ is the average fraction of energy lost to the pion, 
and $\varepsilon_0 = 0.15 \GeV$ is the threshold energy. 
The specific photon density, $n(x)$, is related to the observed luminosity by  
$n(x) = L_\varepsilon(\Gamma x) /4 \pi r^2 c \Gamma x$. 
Photo-meson 
production is dominated by interaction with photons in the energy 
range $\varepsilon_\gamma^{ob.} > \Gamma \varepsilon_0 /2 \gamma_{p,dyn} \approx 
\varepsilon_{\gamma,m,r}^{ob.}$, where  
$L_\varepsilon  \propto\varepsilon_{\gamma}^{-1}$, thus     
\beq 
t_\pi^{-1}  = \frac{2}{3\pi r^2} \frac{L_m \gamma_p}{\Gamma^2}  
\frac{\varepsilon_{\gamma,m,r}^{ob.}}{\varepsilon_{peak}} 
\frac{\sigma_{peak} \xi_{peak} \Delta \varepsilon}{\varepsilon_{peak}}.
\label{t_pi}  
\eeq 
A contribution from the $\Delta$-resonance comparable to that of 
photons of higher energy was assumed in evaluating the first integral. 
Using $\sigma_{peak} \simeq 5 \times 10^{-28} \rm{\, cm^2}$ and  
$\xi_{peak} \simeq 0.2$ at the resonance  
$\varepsilon = \varepsilon_{peak} = 0.3 \GeV$, 
and $\Delta \varepsilon \simeq 0.2 \GeV$ for the peak width, the acceleration time equals the pion production loss time for protons with (plasma frame) Lorentz factor
\beq 
\gamma_{p,\pi} \approx 4.6 \times 10^{9} \, E_{53}^{1/8} T_2^{3/8}
A_*^{-3/8} \epsilon_{e,-1}^{-1/2} \epsilon_{B,-1}^{-1/4}.
\label{eq:gamma_pi} 
\eeq   

Comparing eqs.~(\ref{gamma_p_M}),~(\ref{gamma_syn})
and~(\ref{eq:gamma_pi}) we find that proton acceleration is limited by
energy losses, and that pion production losses dominates over
synchrotron losses for  
\beq 
\epsilon_{e}\gtrsim \epsilon_{B}/10. 
\label{eq:t_syn_pi}
\eeq 
Note, that this is valid independent of the proton energy, since
$t_\pi\propto t_{syn}\propto \gamma_p^{-1}$. X-ray afterglow
observations suggest that $\epsilon_e$ is close to equipartition,
$\epsilon_e \sim 1/3$ \citep{FW01,BKF03}. Thus, we expect proton
energy losses to be dominated by pion production. Comparing
eqs.~(\ref{gamma_p_M}) and~(\ref{eq:gamma_pi}) we find that for
$\epsilon_e\sim0.1$ the highest energy protons lose all their energy
to pion production. Assuming a power law distribution of proton
energies, $dn_p/d\gamma_p \propto\gamma_p^{-2}$, roughly 0.1 of the
energy carried by protons will be converted in this case to pions, and
roughly half of this energy would be converted to high energy photons
through the decay to photons and electrons (positrons). Since shock
accelerated electrons lose all their energy to radiation, this implies
that the contribution of protons to the luminosity is similar to that
of the electrons. This is valid also for values of $\epsilon_e$ well
below equipartition, $\epsilon_e\ll 0.1$, since the contribution to
the luminosity of both electrons and protons is
$\propto\epsilon_e^{-1}$ in this case.  

In the case that eq.~(\ref{eq:t_syn_pi}) is not satisfied, and the
proton energy loss is dominated by synchrotron losses, comparing
eqs.~(\ref{gamma_p_M}) and~(\ref{gamma_syn}) implies that the highest
energy protons lose all their energy to synchrotron losses if the
magnetic field is close to equipartition. For lower values of
$\epsilon_B$, the highest energy protons lose a fraction
$\sim\epsilon_B^{3/2}$ of their energy by synchrotron emission. In
this case, proton synchrotron emission would dominate the electron
luminosity if $\epsilon_B^{3/2}>10\epsilon_e$.  
  
For explosion into a uniform density ISM, similar arguments imply that
pion production losses dominate over synchrotron losses when
eq.~(\ref{eq:t_syn_pi}) is satisfied. 
Following the analysis of \citet{WB00},
protons which are expected to be accelerated up to  $\gamma_{p,dyn}
\approx 5 \times 10^{9} \, E_{53}^{1/4} n_0^{1/4} T_2^{1/4}
\epsilon_{B,-1}^{1/2}$, lose a fraction $f_\pi(\gamma_p) \approx 2 \times
10^{-11} \gamma_p \, E_{53}^{3/8} n_0^{5/8} T_2^{-1/8}
\epsilon_{e,-1}$ of their energy to pion production. 
Therefore, for a power law index $p=2$ of
the accelerated protons, pions receive a fraction $ \simeq 10^{-3} \,  
E_{53}^{5/8} n_0^{7/8} T_2^{1/8} \epsilon_{e,-1}
\epsilon_{B,-1}^{1/2}$ of the total proton energy. Since in this
scenario as well electrons lose almost all their energy to radiation, 
the total proton contribution to the photon flux is $\approx 5 \times
10^{-3}$ of the electron contribution, independent on the value of
$\epsilon_e$. 
If eq.~(\ref{eq:t_syn_pi}) is not satisfied, proton synchrotron losses
dominate over pion production. A similar calculation to the former case
shows that proton synchrotron emission dominates the electron
luminosity in this scenario only if $\epsilon_B^{3/2}>10^4 \epsilon_e$.

The discussion presented above demonstrates that the energy loss of
protons is expected to be dominated by pion production, and that this
energy loss may produce a luminosity similar to that of the electrons
in the scenario of explosion into a wind. Finally, it should be
pointed out that if protons are accelerated to a power-law energy
distribution with an index $p$ significantly larger than 2, their
contribution to the luminosity will be significantly reduced, since
the fraction of energy carried by the highest energy protons will be
small.

\section{The numerical model} 
\label{sec:numerics} 

The acceleration of particles in the two shock waves that exist during
the transition phase is accompanied by numerous radiative processes.
In the numerical calculations, we use the time dependent
  numerical model described in \citet{PW03}. 
Our time dependent model follows the evolution of the particle
distribution and the emergent spectra, by solving the kinetic
equations for the electron and photon distributions, taking into
account synchrotron emission, inverse Compton scattering, pair
production and pion photo-production interactions.  
These calculations are done for a wide range of particle energies, 
including the evolution of rapid electro-magnetic cascades at high
energies. 

In the calculations, we focus on particles that pass through the
forward or the reverse shock waves and therefore are in the downstream
region of the flow, relative to the relevant shock wave. 
While the upstream relativistic flow is highly anisotropic, the
shocked gas thermalize, hence isotropise (in the plasma comoving
frame) immediately after passing the shock, on a characteristic length
scale of several skin-depth \citep[e.g.,][]{Kirk98, Kirk00, FHHN04}. The emitted
radiation is therefore isotropic in the comoving frame.
Most of the shocked gas and most of the blast-wave energy are
concentrated in a shell of co-moving thickness $\Delta r = \zeta c
T_{dyn}$  with $\zeta \lesssim 1$ (e.g., in the \citet{BM76}
self-similar solutions which give the spatial dependence of the
hydrodynamical variables, 90\% of the energy is concentrated in a
shell of co-moving thickness corresponding to $\zeta = 1/7 $).
Since the details of the spatial dependence of the electron and
magnetic field energy fractions are not known, we adopt in our model
the commonly used approximation \citep[e.g.,][]{LZ87,Coppi92, PL98} that
radiation is produced within a homogeneous shell of comoving width
$\Delta r =\zeta c T_{dyn}$ with $\zeta = 1$. Since the shocks
velocity is time independent during $T_{dyn}$, the shock-heated
comoving plasma volume is assumed to increase linearly with time,
i.e., constant particle number density is assumed \citep[see][for
 further details]{PW03}.

In the comoving frame, homogeneous and isotropic distributions of both
particles and photons are therefore assumed. 
Parallel calculations of processes in the two shock waves are carried
out, where photons produced at each shock participate in IC scattering,
photo-production interactions and pair production interactions
occurring at both shocks.

The particle distributions are discretized, the proton spectrum
spans 11 decades of energy ($\gamma_p < 10^{11}$), and the electron
spectrum 14 decades ($\gamma_e < 10^{14}$). A fixed time step is
chosen, typically $10^{-4}$ times the dynamical time. Numerical
integration is carried out with this fixed time step. Particles and
photons for which the energy loss time or annihilation time are
shorter than the fixed time step, are assumed to lose all their
energy in a single time step, producing secondaries which are treated
as a source of lower energy particles. Photo-pion production is
calculated at each time step by direct integration of the second
integral in Eq.~\ref{eq:t_pi_int}, while approximating the first
integral by the contribution from the $\Delta$-resonance (see the
discussion following Eq.~\ref{t_pi}). Half of the energy lost by
protons goes into $\pi^0$ that decays into 2 photons, each carrying 
$10\%$ of the initial proton energy.  

The following approximations are made: 
(i) Plasma parameters are assumed to be time independent during the
transition phase;  
(ii) The fraction of thermal energy  that is carried by electrons
(magnetic field), $\epsilon_e$  ($\epsilon_B$), is the same at the
forward and reverse shock waves; 
(iii) The power law index of the energy distribution of accelerated
electrons and protons is the same at both shocks. 

In the present calculations, we do not consider
proton synchrotron emission, which is dominated by 
pion production for the considered parameter range.
Synchrotron self absorption is also not considered here, being 
irrelevant for processes occurring at high energies. Photons below the
self absorption frequency, $\varepsilon_{ssa}^{ob.} \leq 1 \eV$ 
(in both scenarios; see eqs. \ref{eq:ssa_wind}, \ref{eq:ssa_ism}) can
produce pairs only with photons energetic than $\Gamma^2 (m_e c^2)^2 /
\varepsilon_{ssa}^{ob.} \simeq 10^{16} \eV$, which are absent (see the
numerical results below).
Moreover, these photons are below the threshold energy for pion
production, $ \varepsilon^{ob.}_{TH.} \geq 0.2 \Gamma \GeV / \gamma_p
= 10 \eV$, where $\gamma_{p,\pi} = 5 \times 10^9$ (see
eq.~\ref{eq:gamma_pi}), and $\Gamma  =10^{2.5}$ where taken.

\section{Numerical results} 
\label{sec:results} 
 
\subsection{High vs. low density} 

\begin{figure} 
 \plotone{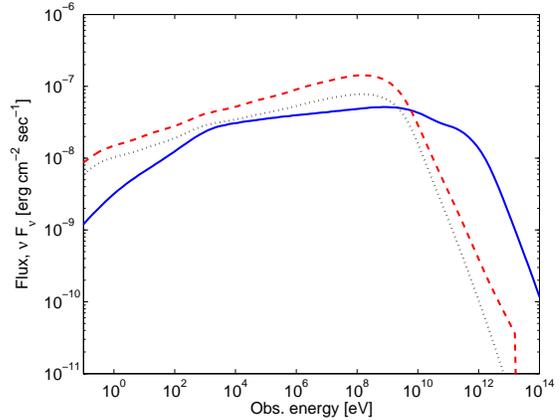} 
\caption{Predicted photon spectrum during the transition phase. 
   Results are shown for  $E= 3\times10^{53}$~erg , $\Gamma_i = 10^{2.5}$,
   $T = 10 \rm\,{s}$,   $\epsilon_e = 10^{-1}$, $\epsilon_B =
   10^{-2}$, $p=2.0$.
   Solid: explosion into uniform low density medium (ISM), 
   $n =  1 \rm{\,cm^{-3}}$. Dashed: Explosion into a wind with $A_* =
   1 \rm{\, gr \,cm^{-1}}$. Dotted: Explosion into a wind with $A_* =
   1 \rm{\, gr \,cm^{-1}}$, proton contribution to the flux omitted. 
   Luminosity distance $d_L = 2 \times 10^{28} \rm{\, cm}$
   and $z=1$ were assumed. Intergalactic absorption of high energy
   photons through pair production interactions with the IR
   background, which becomes appreciable above $\sim0.1$~TeV, is not
   taken into account. 
\label {fig_result}} 
\end{figure}

Figure \ref{fig_result} shows spectra obtained from a GRB transition phase
in the two scenarios discussed: Explosion in a uniform low density
medium typical to the ISM, $n = 1 \rm{\, cm^{-3}}$ , and explosion in
a $A_*=1$ wind, where the density of ambient medium at the transition
radius is much higher, $n \approx 10^3 \rm{\, cm^{-3}}$. 

The most evident feature of the two spectra is their flatness, $\nu
F_\nu \propto \nu^{\alpha}$ with $\alpha \simeq 0$, extending over
seven energy decades between $\varepsilon_{\gamma,m,f}^{ob.} \simeq
1.4 \keV$ and $\varepsilon_{\gamma,\max}^{ob.} \simeq 60 \GeV$ for
expansion into ISM, and over ten energy decades from
$\varepsilon_{\gamma,m,r}^{ob.} \approx 1 \eV$ to
$\varepsilon_{\gamma,\max}^{ob.} \approx 10 \GeV$ for explosion into a
wind. This is a result of dominant synchrotron emission term at these energies.

\begin{figure}[ht] 
  \plotone{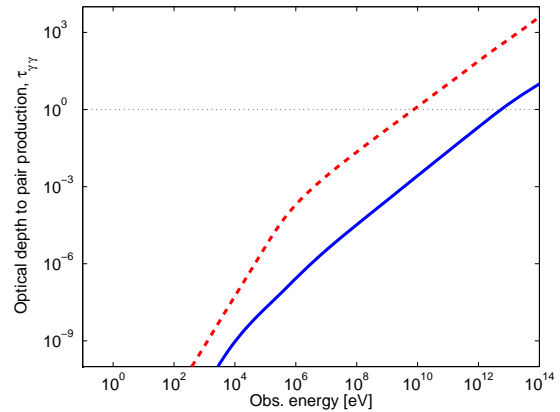} 
  \caption{Energy dependent optical depth to pair production,
    for the two scenarios considered in figure \ref{fig_result}.
    Solid: explosion into ISM, dashed: explosion into a wind.
    All physical parameters are the same as in figure  \ref{fig_result}. 
\label {fig_tao}} 
\end{figure}

\begin{figure}[ht] 
  \plottwo{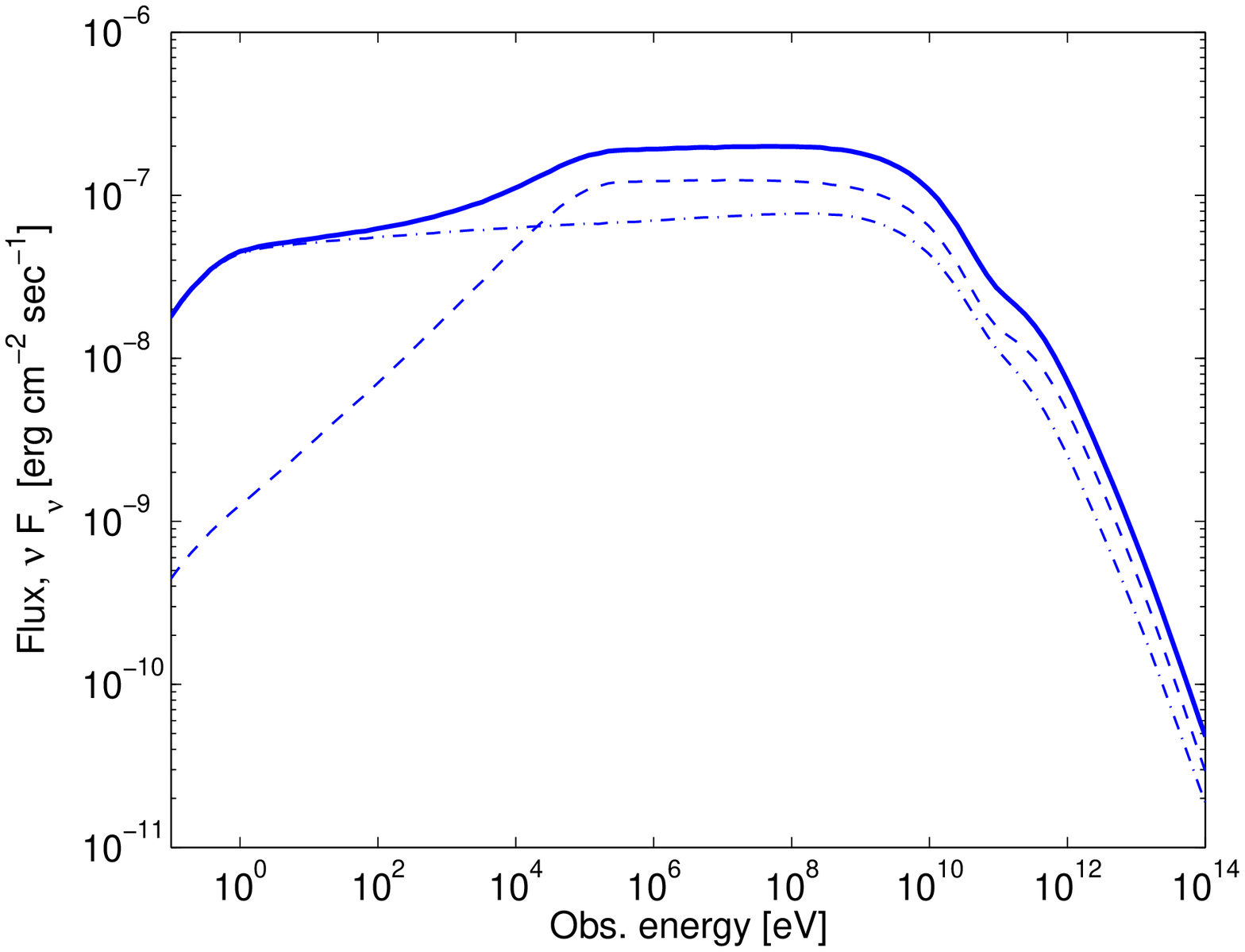}{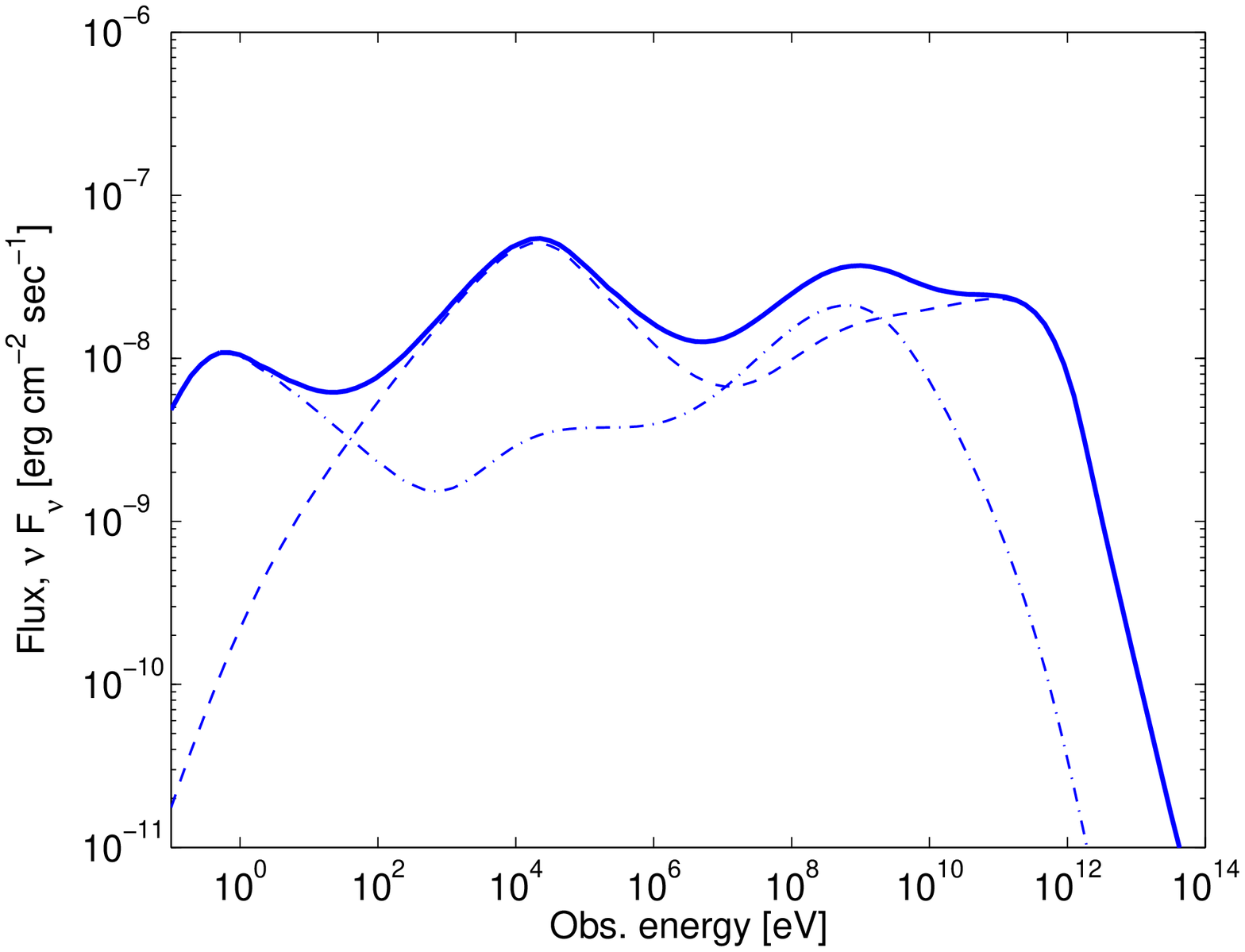}
  \caption{
Two emission components of the observed spectra, explosion into ISM.
Dash: forward shock emission, dash-dotted: reverse shock emission,
solid: combined spectra.
Left: equipartition, $\epsilon_e = 10^{-0.5}$, $\epsilon_B = 10^{-0.5}$, $p=2.0$.
The two emission components are comparable.
Right: $\epsilon_e = 10^{-1}$, $\epsilon_B = 10^{-2}$, $p=3.0$.
The wavy shape results from different peaks of the two synchrotron
components, and the two IC components.
All other physical parameters are the same as in figure \ref{fig_result}.
\label{fig_components}} 
\end{figure}

The main difference between the two scenarios appears at high energies, 
$100 \GeV - 1 \TeV$, due to the different optical depth to pair production, 
$\tau_{\gamma \gamma}(\varepsilon) $, as shown in Figure \ref{fig_tao}. 
The large optical depth to pair production in the wind case softens
the high energy IC spectrum, resulting in spectral index $\alpha =
-1$. Consequently, while in the ISM case the $1 \TeV$ flux is
comparable to the flux at $1 \GeV$, in the high density wind case the
$1 \TeV$ flux is 2.5 orders of magnitude lower. A second difference
between the two scenarios appears at the optical-UV band. 
In the low density scenario, at low energies, 
$\varepsilon \lesssim \varepsilon_{\gamma,m,f}^{ob.}$,
emission from both shock waves produces a  moderate increase in the
spectral slope,  $\alpha \simeq 0.3$ above 
$\varepsilon_{\gamma,c}^{ob.} \approx 20 \eV$.
As demonstrated in Figure \ref{fig_components}, this slope results
from a flat spectrum produced at the reverse shock, $\alpha = 0$ above 
 $\varepsilon_{\gamma,m,r}^{ob.} \approx 10^{-2} \eV$,
and forward shock emission characterized by 
$\alpha =0.5 $ below $\varepsilon_{\gamma,m,f}^{ob.}$.
In a high density medium, a nearly flat spectrum is obtained above
$\varepsilon_{\gamma,m,r}^{ob.} \approx 1 \eV$. 

In an explosion into constant density ISM scenario, the total energy
 carried by the IC scattered photons is $\approx 10\%$ of the total
 energy in synchrotron emitted photons, well below the simple estimate
 $(\epsilon_e/\epsilon_B)^{1/2}$, which is based on the assumption
 that all the synchrotron emitted photons serve as seed photons for
 Compton scattering. 
This discrepancy is due to Klein-Nishina suppression of the
 cross-section at high energies. 
Consequently, IC scattering becomes dominant only  
above $\sim 50 \GeV$, where a flat spectrum is created  
for a power law index $p=2$ of the accelerated electrons energy
 distribution. In the wind case, IC scattering is more prominent due
 to the higher photon density.    

\begin{figure}[ht] 
  \plotone{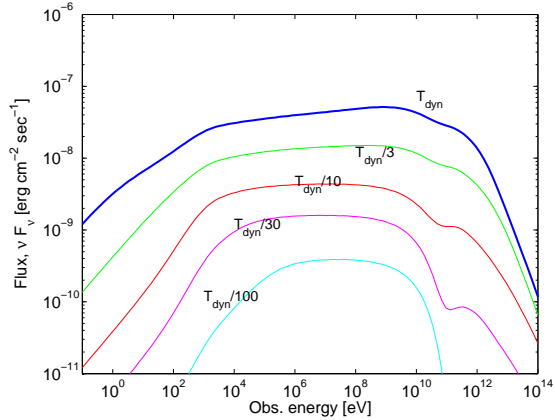} 
  \caption{Lightcurve of the spectra for the scenario of explosion
    into uniform low density medium (same parameters as in figure
    \ref{fig_result}). 
    The lines show the spectrum after fractions of 1/100, 1/30, 1/10,
    1/3 and 1 dynamical time.
\label {fig_lightcurve}} 
\end{figure}

Lightcurve of the spectrum of an explosion into constant density ISM
 scenario is presented in Figure~\ref{fig_lightcurve}. 
The constant injection rate of energetic electrons results in a linear
increase of the synchrotron emission component in time. Since the
energy of photons emitted by electrons for which the synchrotron
cooling time equals the time $t$, $\varepsilon_{\gamma,c}(t) \propto
t^{-2}$, the energy density in the low energy photons increases with
time. Only photons with $\varepsilon_\gamma^{ob.} \lesssim 1 \keV$ are
in the Thompson limit for Compton scattering, therefore the inverse
Compton component increases  in time, at early times. 
At later times, $t \lesssim T_{dyn}$, where
 $T_{dyn}=4\Gamma_{\rm BM} T$ is the dynamical time of the problem,
 the low energy photon density becomes high enough that pair production
 phenomenon limits the high energy emission component.

The contribution of protons to the photon flux is negligible in the
ISM case, however this contribution is 85\% of the electrons'
contribution to the flux in the wind scenario (see \S\ref{sec:prot}).
Most of this contribution results from the decay of energetic pions, 
initiating high energy electro-magnetic cascades. 
The high energy cascade results in flat ($\alpha = 0$) spectrum at the
observed energy range, which, for power law index of the accelerated
particles $p=2$ is similar to the spectrum from inverse Compton
scattering. An additional small  contribution from the decay of low
energy pions results in a spectral slope $\alpha=1$, contributing to
the small deviation from flat spectrum shown below $1 \GeV$ in this
scenario. Above the energy at which the optical depth for pair
production $\tau_{\gamma \gamma} > 1$, $\sim 1 \GeV$, the spectrum is determined
by the spectrum of low energy photons, thus the proton contribution is
hard to discriminate at these energies as well.

\subsection{Fraction of energy carried by electrons and magnetic field} 
\label{sec:B}

\begin{figure}[ht] 
  \plotone{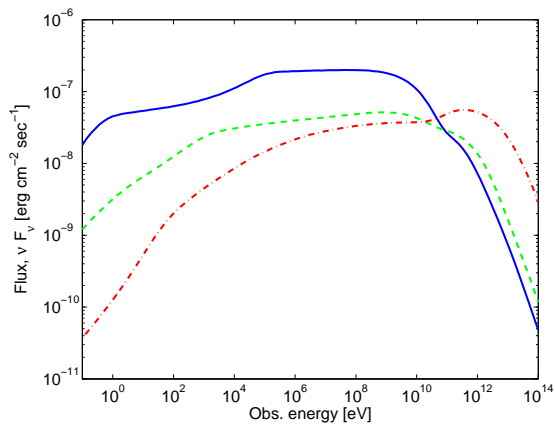} 
  \caption{Dependence of the spectra on the fraction of thermal energy carried by
    the electrons and by the magnetic field, $\epsilon_e$ and $\epsilon_B$,
 explosion into a low density ISM, $n=1 \rm{\, cm^{-3}}$.
Solid: $\epsilon_e = 10^{-0.5}$, $\epsilon_B = 10^{-0.5}$,
dashed:   $\epsilon_e = 10^{-1}$, $\epsilon_B = 10^{-2}$, dash-dotted:
    $\epsilon_e = 10^{-1}$, $\epsilon_B = 10^{-4}$.
All other physical parameters are the same as in figure \ref{fig_result}.
\label {fig_nw_xi_B}} 
\end{figure}

In an explosion into a low density ISM, $\gamma_{m,r} < \gamma_c \ll
\gamma_{\max}$. Thus, for $p \simeq 2$ electrons lose nearly 100\% of
their energy to radiation, resulting in a nearly linear dependence of
the flux on $\epsilon_e$. The dependence of the spectrum on the
fraction of thermal energy carried by the magnetic field, 
$\epsilon_B$, is illustrated in Figure \ref{fig_nw_xi_B}. The
domination of synchrotron emission implies that the values of
$\varepsilon_{\gamma,c}$ and $\varepsilon_{\gamma,m}$ 
determine the spectral behavior at low energies, 
while a flat spectrum is expected for $p=2$ at higher energies, 
up to $\varepsilon_{\gamma,\max} \approx 50 \GeV$.
The ratio of $1 \GeV$ to $1 \keV$ flux is informative
about the value of $\epsilon_B$, with large ratio ($\sim 30$) implying
small $\epsilon_B (\approx 10^{-4}$). At higher energies, $\varepsilon
> 100 \GeV$, a flux comparable to the $1 \GeV$ flux is produced by IC
scattering for $\epsilon_B<0.1$.  

\begin{figure}[ht] 
  \plotone{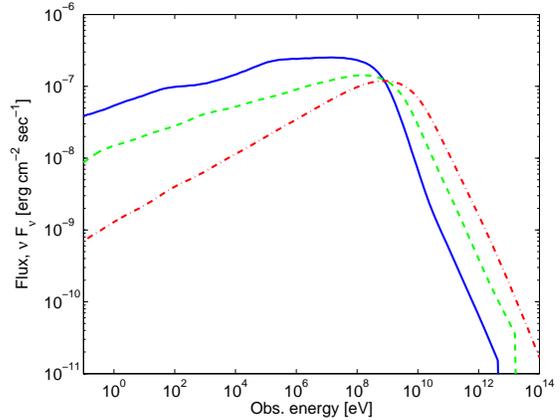} 
  \caption{Dependence of the spectra on the fraction of thermal energy carried by
    the electrons and by the magnetic field, $\epsilon_e$ and $\epsilon_B$,
 explosion into a wind characterized by  $A_*=1 \rm{\, gr \,cm^{-1}}$.
Solid: $\epsilon_e = 10^{-0.5}$, $\epsilon_B = 10^{-0.5}$,
dashed:   $\epsilon_e = 10^{-1}$, $\epsilon_B = 10^{-2}$, dash-dotted:
    $\epsilon_e = 10^{-1}$, $\epsilon_B = 10^{-4}$.
All other physical parameters are the same as in figure \ref{fig_result}.
\label {fig_w_xi_B}} 
\end{figure}

For explosion into a wind, $\gamma_c < \gamma_{m,r} \ll
\gamma_{\max}$. Here too, for $p \simeq 2$ electrons lose nearly 100\%
of their energy to radiation, resulting in a nearly linear dependence
of flux on $\epsilon_e$. The dependence of the spectrum on
$\epsilon_B$ is illustrated in Figure \ref{fig_w_xi_B}. Here, IC
emission is much more prominent. IC energy loss of the electrons
affects the spectrum at all energy bands by modifying the electrons
energy distribution. The spectrum becomes harder with lower values of
$\epsilon_B$, reaching $\alpha \sim 0.2$ below $10 \GeV$ for
$\epsilon_B = 10^{-4}$. 
In this scenario, the minimum Lorentz factor of electrons accelerated
at the front shock, $\gamma_{\min,f} = 2.0 \times 10^3$ (see Eq.~
\ref{eq:gamma_min}) is comparable to $\gamma_{IC-S} \approx 10^4$,
which is defined as the Lorentz factor of electrons that emit
synchrotron radiation above their Klein-Nishina limit.  
We show in Appendix \ref{sec:sync_ic} that in this case, if the
main energy loss mechanism is inverse Compton scattering, a spectral
index in the range $0\leq \alpha \leq 2/3$ should be expected. A $1
\GeV$ to $1 \keV$ flux ratio larger than 10 indicates therefore
$\epsilon_B \approx 10^{-4}$, and smaller ratio indicates higher value
of $\epsilon_B$, similar to the case of explosion into low density
ISM. Above $\sim 10 \GeV$, the flux is suppressed by pair production.

\subsection{Power law index of the accelerated particles} 

\begin{figure}[ht] 
  \plotone{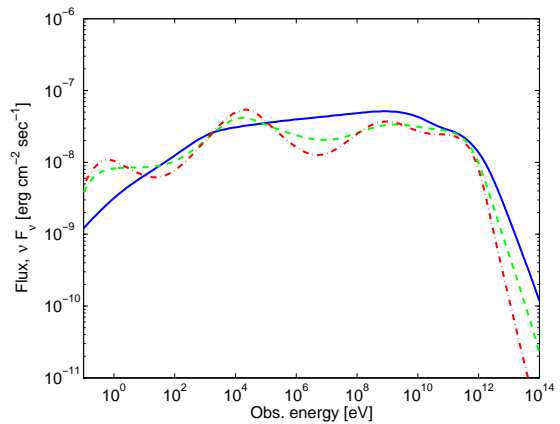} 
  \caption{Dependence of the spectra on the power law index $p$ of the
    accelerated electrons, explosion into low density ISM, $n=1 \rm{\, cm^{-3}}$.
Solid: $p=2.0$, dashed: $p=2.5$, dash-dotted: $p=3.0$.
All other physical parameters are the same as in figure \ref{fig_result}.
The wavy shape of the spectrum results from synchrotron and IC
components of the two shock waves, as seen in figure \ref{fig_components}. 
\label {fig_nw_power_law}} 
\end{figure}

\begin{figure}[ht] 
  \plotone{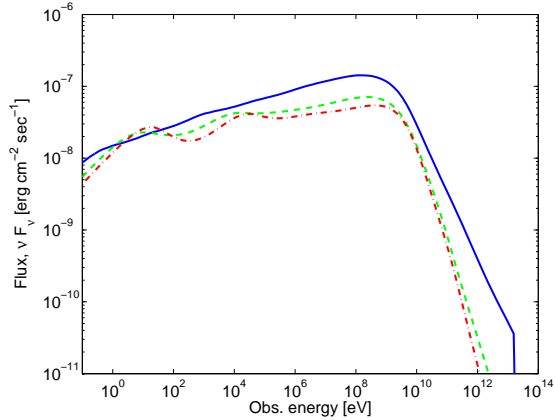} 
  \caption{Dependence of the spectra on the power law index $p$ of the
    accelerated electrons, explosion into a wind with $A_*=1 \rm{\, gr \,cm^{-1}}$.
Solid: $p=2.0$, dashed: $p=2.5$, dash-dotted: $p=3.0$.
All other physical parameters are the same as in figure \ref{fig_result}.
\label {fig_w_pow_low}} 
\end{figure}

The dependence of the spectrum on $p$ varies with the external density, as
demonstrated in Figures \ref{fig_nw_power_law} and \ref{fig_w_pow_low}.
The dominant synchrotron component in explosion into a low density ISM 
results, for large $p$, in a complex wavy shape shown in Figure
\ref{fig_nw_power_law}. The four peaks result from synchrotron and IC
components at the two shock waves, as demonstrated in Figure
\ref{fig_components} (right). The dependence on $p$ is, however, not
prominent. For explosion into a wind, the prominent IC component leads
to a spectrum which is in nearly independent on $p$.

\section{Discussion} 
\label{sec:discussion} 

We have presented numerical results of calculations of the early
afterglow GRB spectra (figures \ref{fig_result},
\ref{fig_nw_xi_B}-\ref{fig_w_pow_low}) within the fireball model
framework. 
Our time dependent numerical code describes synchrotron emission,
inverse-Compton scattering, $e^\pm$ pair production, photo-pion
production and the resulting high energy cascades.
We have shown that the dependence of the spectra on the ambient density and
the magnetic field equipartition fraction is pronounced mainly at high
photon energies, $1 \GeV - 1\TeV$ (see Figures  \ref{fig_nw_xi_B}, \ref
{fig_w_xi_B}): Comparable flux at 1~GeV and 1~TeV 
implies low density and  $\epsilon_B \simeq 10^{-4}$,
while a large ratio of $\sim 10^3$ results from dense medium and
$\epsilon_B$ near equipartition.
A second indication for $\epsilon_B \simeq 10^{-4}$ is a large 
($\sim 30$) ratio of the flux at $1 \GeV$ to the flux at $1 \keV$.
In all cases, the spectra depend only weakly on the power-law index
$p$ of the accelerated electrons. The flux expected in the $1-50 \GeV$ energy band, 
$\nu F_\nu \sim 10^{-8} -10^{-7} \flu$ for $z=1$ bursts, is well within the
detection capability of GLAST. 
  
The recent detection by MILAGRITO of $1 \TeV$ photon flux associated
with GRB970417a \citep{Atkins03} can be explained as due to the onset
of fireball deceleration by a medium with density typical to the ISM,
provided that the magnetic field is well below equipartition,
$\epsilon_B \lesssim 10^{-4}$. The flux at $1 \TeV$ is expected in
this case to be
an order of magnitude higher than the $100 \keV$ (BATSE) flux, as
inferred from the MILAGRITO detection. We note, that an alternative
explanation was suggested for the $1\TeV$ flux, synchrotron emission
from shock accelerated protons \citep{Totani98a,Totani98b}. This
explanation requires a very low fraction of the energy to be carried
by electrons, $\epsilon_e \sim 10^{-3}$, and an isotropic equivalent
explosion energy $E\sim10^{56} \rm{\, erg}$. We find this explanation
less attractive, since X-ray afterglow observations imply $\epsilon_e$
values near equipartition and isotropic equivalent energies in the
range of $\sim 10^{51.5}$~erg to $10^{53.5}$~erg \citep{FW01, BKF03}.  

The energy loss of high energy protons is dominated by
photo-production of pions, rather than by synchrotron emission, for
$\epsilon_e>\epsilon_B/10$. The contribution of pion decay to the high
energy photon luminosity may be comparable to that of inverse-Compton
emission of shock accelerated electrons in the case of fireball
expansion into a high density wind, and power law index $p\simeq2$ of
the accelerated particles. However, it is difficult to
distinguish in this case between the electron and proton contributions
since the spectral shape is determined primarily by the energy
dependence of the pair production optical depth. 
In the two scenarios considered, explosion into constant density ISM
and into a pre-ejected wind, if $\epsilon_e \leq 10^{-5} - 10^{-4}$, proton
synchrotron emission becomes the dominant emission mechanism.

Although the MILAGRITO detection is singular, the new generation of high
energy sub-TeV detectors, such as MILAGRO,  
 VERITAS, 
HEGRA, 
HESS,
and MAGIC 
that will become fully operational in the next few years, will be
sensitive enough to observe the high energy early afterglow emission
at $100 \GeV$ from $\sim 10$ bursts per year. The detection rate of
higher energy, $1 \TeV$ component is much lower, due to the large pair production optical depth for TeV photons originating at $z>0.3$ \citep[e.g.,][]{SS98}. However, the predicted $1 \TeV$ flux implies
that $\approx 1$ burst per year, that arrives from low redshift $z
\leq 0.3$, could be detected in the $1 \TeV$ band as well. Such
detections will allow to constrain two of the uncertain parameters of the fireball model: the
ambient gas density, and the magnetic field equipartition fraction. 

\acknowledgements
This work was supported in part by a Minerva grant and an ISF grant. 
AP wishes to thank Amir Sagiv for valuable discussions.

\appendix

\section{Spectral distribution resulting from 
Synchrotron radiation and IC scattering} 
\label{sec:sync_ic} 
If inverse Compton scattering is the dominant energy loss mechanism of
electrons that were accelerated with a power law energy distribution $d n_e /
d\gamma \propto \gamma^{-p}$, the resulting power law index
of energetic electrons above $\max(\gamma_m , \gamma_c)$ deviates from $p+1$.  

The resulting power law can be calculated analytically if the minimum
Lorentz factor of the accelerated electrons $\gamma_m$, is comparable
to $\gamma_{IC-S}$, which is defined to be the Lorentz factor of
electrons that emit synchrotron radiation at the Klein-Nishina
limit, $\hbar (3 q B \gamma_{IC_S}^2)/(2 m_e c) = 3/4 (m_ec^2)/
\gamma_{IC_S}$. At steady state, a broken power law is expected
with index $(p_1+1)$ above $\gamma_{IC-S}$ and $(p_2+1)$ at lower energies.
Synchrotron emission has a broken power law spectrum: 
\begin{eqnarray} && 
\varepsilon dn_\gamma/d\varepsilon \propto \varepsilon^{-p_1/2} 
\rightarrow U_\gamma \propto \varepsilon^{1-p_1/2} 
\qquad (\varepsilon>\varepsilon_{IC-S}), \nonumber \\&& 
\varepsilon dn_\gamma/d\varepsilon \propto \varepsilon^{-p_2/2} 
\rightarrow U_\gamma \propto \varepsilon^{1-p_2/2} 
\qquad (\varepsilon<\varepsilon_{IC-S}), 
\end{eqnarray} 
where $\varepsilon_{IC-S} \equiv \hbar (3 q B \gamma_{IC_S}^2)/(2 m_e c)$.
Energetic electrons therefore lose energy at a rate 
$d\gamma/dt \propto \gamma^2 U_\gamma \propto \gamma^{1+p_2/2}$. 
As these electrons are injected with power law index $p$,
their steady distribution is 
$dn_e(\gamma>\gamma_{IC-S})/d\gamma \propto \gamma^{-(p+p_2/2)}$, or
\beq 
p_1 = p + \frac{p_2}{2} -1. 
\label{p1} 
\eeq 
At lower energies, electrons lose energy by scattering photons above  
 $\varepsilon_{IC-S}$, thus $dn_e(\gamma<\gamma_{IC-S})/d\gamma
\propto \gamma^{-(1+p_1/2)}$, or  
\beq 
p_2 = \frac{p_1}{2}. 
\label{p2} 
\eeq
Combined together, one obtains
\begin{eqnarray}&& 
p_1 = \frac{4}{3} (p-1), \nonumber \\&& 
p_2 = \frac{2}{3} (p-1). 
\label{eq:p_1_2}
\end{eqnarray} 
 As an example, assuming power law index $p=2$ of the accelerated
 electrons, the photons spectral index is 1/3 at low energies, or 
$\nu F_\nu \propto \nu^{2/3}$. 

Note that the above analysis is valid provided both $p_1, p_2 < 2$,
 which is translated via eq.~(\ref{eq:p_1_2}) into the demand $p<2.5$.


\begin{thebibliography}{} 
 
\bibitem [Akerlof {\it et al.} (1999)]{Akerlof99} 
  Akerlof, C.W., {\it et al.} 1999, \nat, 398, 400 
 
\bibitem [Atkins {\it et al.} (2000)]{Atkins00} 
  Atkins, R., {\it et al.} 2000, \apj, 533, L119 
 
\bibitem [Atkins {\it et al.} (2003)]{Atkins03} 
  Atkins, R., {\it et al.} 2003, \apj, 583, 824 
 
\bibitem[Berger, Kulkarni \& Frail (2003)]{BKF03}
Berger, E., Kulkarni, S.R., \& Frail, D.A. 2003, \apj, 590, 379
 
\bibitem[Blandford \& McKee (1976)]{BM76} 
  Blandford, R.D., \& McKee, C.F. 1976, Phys. Fluids 19, 1130 
 
\bibitem[Bonometto \& Rees (1971)]{BR71}
 Bonometto, S., \& Rees, M.J. 1971, \mnras, 152, 21

\bibitem[B\"otcher \& Dermer (1998)]{BD98} 
  B\"otcher, M. \& Dermer, C.D. 1998, \apj, 499, L131 
 
\bibitem[Chiang \& Dermer (1999)]{CD99} 
  Chiang, J., \& Dermer, C.D. 1999, \apj, 512, 699 
 
\bibitem[Chevalier (2001)]{Chevalier01} 
  Chevalier, R.A., 2001, in GRB in the afterglow era, Proceedings of the International workshop held in Rome, CNR (astro-ph/0102212) 
 

\bibitem[Coppi (1992)]{Coppi92}
 Coppi, P.S. 1992, \mnras, 258, 657

\bibitem [De Paolis, Ingresso \& Orlando (2000)]{PIO00} 
  De Paolis, F., Ingresso, G., \& Orlando, D. 2000, \aap, 359, 514 
 
\bibitem [Eichler {\it et al.} (1989)]{Eichler89} 
  Eichler, D., Livio, M., Piran, T., \& Schramm, D.N. 1989, \nat, 340,
  126 

\bibitem [Fabian {\it et al.} (1986)]{FBGPC}
 Fabian, A.C., Blandford, R.D., Guilbert, P.W., Phinney, E.S., \& Cuellar, L.
1986, \mnras, 221, 931
 

\bibitem [Frederiksen  {\it et al.}  (2004)]{FHHN04}
  Frederiksen, J.T., Hededal, C.B., Haugb\o lle, T. \& Nordlund,
  \AA. 2004, \apj, 608, L13

\bibitem [Freedman \& Waxman (2001)]{FW01}
  Freedman, D.L., \& Waxman, E. 2001, \apj, 547, 922

\bibitem[Goodman (1986)]{Good86} 
  Goodman, J. 1986, \apj, 308, L47 
 
\bibitem [Gruzinov \& Waxman (1999)]{GW99} 
  Gruzinov, A. \& Waxman, E. 1999, \apj, 511, 852 
  
\bibitem[Guetta, Piran \& Waxman (2003)]{GPW03}
  Guetta, D., Piran, T.  \& Waxman, E. 2003, \apj, 619, 412
 
\bibitem [Guilbert, Fabian \& Rees (1983)]{Guilbert83}
 Guilbert, P.W., Fabian, A.c., \& Rees, M.J. 1983 \mnras, 205, 593

\bibitem[Halzen (1999)]{Halzen99} 
  Halzen, F. 1999, in Proc. 17th Int. Workshop 
on weak interactions and Neutrinos, 
Cape Town, South Africa (astro-ph/9904216) 
 
\bibitem [Hurley {\it et al.} (1994)]{Hurley94} 
  Hurley, K. {\it et al.} 1994, \nat, 372, 652 
  
\bibitem[Katz (1994)]{Katz94} 
  Katz, J.I. 1994, \apj, 432, L27 
 
\bibitem[Kirk {\it et al.} (1998)]{Kirk98}
  Kirk, J.G., Rieger, F.M., \&  Mastichiadis, A. 1998, A\&A, 333, 452
 
\bibitem[Kirk {\it et al.} (2000)]{Kirk00}
  Kirk, J.G., Guthmann, A.W., Gallant, Y.A., \& Achterberg, A. 2000,
  \apj, 542, 235

\bibitem[Kobayashi (2000)]{Kob00}
  Kobayashi, S. 2000, \apj, 545, 807

\bibitem[Kobayashi \& Sari (2000)]{KS00}
  Kobayashi, S., \& Sari, R. 2000, \apj, 542, 819

\bibitem[Levinson \& Eichler (1993)]{Levinson93} 
  Levinson, A., \& Eichler, D. 1993, \apj, 418, 386 

\bibitem [Lightman \& Zdziarski (1987)]{LZ87}
 Lightman, A.P., \& Zdziarski, A.A. 1987, \apj, 319, 643

\bibitem[M\'esz\'aros (2002)]{fireballs2}
  M\'esz\'aros, P. 2002, ARA\&A 40, 137

\bibitem[M\'esz\'aros \& Rees (1997)]{MR97} 
  M\'esz\'aros, P., \& Rees, M.J. 1997, \apj, 476, 232 
 
\bibitem[M\'esz\'aros \& Rees (1999)]{MR99} 
  M\'esz\'aros, P., \& Rees, M.J. 1999, \mnras, 306, L39 
 
\bibitem[M\'esz\'aros, Rees \& Papathanassiou(1994)]{MRP94} 
  M\'esz\'aros, P., Rees, M.J., \& Papathanassiou, H. 1994, \apj, 432, 181 

\bibitem[Milgrom \& Usov (1995)]{MnU95}
Milgrom, M., \& Usov, V. 1995,  \apjl, 449, L37

\bibitem[Paczy\'nski (1986)]{Pac86} 
  Paczy\'nski, B. 1986, \apj, 308, L43 
 
\bibitem[Paczy\'nski (1998)]{Pac98} 
  Paczy\'nski, B. 1998, \apj, 494, L45 
 
\bibitem[Paczy\'nski \& Rhoads (1993)]{PR93} 
 Paczy\'nski, B., \& Rhoads, J.E. 1993, \apj, 418, L5 
 
\bibitem[Pe'er \& Waxman (2004)]{PW03} 
 Pe'er, A., \& Waxman, E. 2005, \apj, in press (astro-ph/0409539) 
 
\bibitem[Panaitescu \& M\'esz\'aros (1998)]{Pan98} 
  Panaitescu, A., \& M\'esz\'aros, P. 1998 \apj, 501, 702 

\bibitem [Pilla \& Loeb (1998)]{PL98}
  Pilla, R.P., \& Loeb, A. 1998, \apj, 494, L167

\bibitem[Piran (2000)]{fireballs1}
  Piran, T. 2000, Phys. Rep. 333, 529.

\bibitem[Salamon \& Stecker (1998)]{SS98} 
  Salamon, M. H., \& Stecker, F.W. 1998, \apj, 493, 547  
 
\bibitem[Sari \& Piran (1995)]{SP95} 
  Sari, R., \& Piran, T. 1995, \apj, 455, L143  
 
\bibitem[Sari \& Piran (1999)]{SP99} 
  Sari, R., \& Piran, T. 1995, \apj, 517, L109  
 
\bibitem[Sari, Piran \& Narayan (1998)]{Sari98} 
  Sari, R., Piran, T., \& Narayan, R. 1998, \apj, 497, L17  

\bibitem[Schneid {\it et al.} (1992)]{Schneid92} 
  Schneid, E.J. {\it et al.} 1992, \aap, 255, L13 
 
\bibitem[Svensson (1987)]{Svensson87}
 Svensson, R. 1987, \mnras, 227, 403

\bibitem[Totani (1998a)]{Totani98a} 
  Totani, T. 1998, \apj, 502, L13 
 
\bibitem[Totani (1998b)]{Totani98b} 
  Totani, T. 1998, \apj, 509, L81 
 
\bibitem[Vietri (1995)]{Viet95} 
  Vietri, M. 1995, \apj, 453, 883 
 
\bibitem[Vietri (1997a)]{Viet97} 
  Vietri, M. 1997, \apj, 478, L9 
 
\bibitem[Vietri (1997b)]{V97} 
  Vietri, M. 1997, \prl, 78, 4328 
 
\bibitem[Wang, Dai \& Lu (2001)]{Wang01} 
  Wang, X.Y., Dai, Z.G. \& Lu, T. 2001, \apj, 556, 1010 
 
\bibitem[Waxman (1995a)]{W95} 
  Waxman, E. 1995a, \prl, 75, 386 
 
\bibitem[Waxman (1995b)]{w95b} 
  Waxman, E. 1995b, \apj, 452, L1 

\bibitem[Waxman (1997a)]{W97a}
Waxman, E. 1997,  \apjl, 485, L5 

\bibitem[Waxman(1997b)]{W97b}
Waxman. E. 1997,  \apjl, 489, L33

\bibitem[Waxman (1997c)]{W97} 
  Waxman, E. 1997, \apj, 491, L19 
 
\bibitem[Waxman (2003)]{W03} 
  Waxman, E. 2003, in Supernovae and Gamma-Ray Bursts, Ed. K. Weiler (Springer), 
Lecture Notes in Physics 598, 393--418  (astro-ph/0303517) 
 
\bibitem [Waxman \& Bahcall (1997)]{WB97} 
  Waxman, E., \& Bahcall, J. 1997, \prl, 78, 2292 
 
\bibitem[Waxman \& Bahcall (2000)]{WB00} 
  Waxman, E., \& Bahcall, J. 2000, \apj, 541, 707 
 
\bibitem[Waxman \& Draine (2000)]{WD00} 
  Waxman, E., \& Draine, B.T. 2000, \apj, 537, 796 
 
\bibitem[Wei (2003)]{Wei03} 
  Wei, D.M. 2003, \aap, 402, L9 
 
\bibitem[Willis (1991)]{Willis91} 
  Willis, A.J. 1991, in Wolf-Rayet Stars and Interrelations with other Massive stars in Galaxies, ed. K.A. Van der Hucht \& B. Hidayat (Dordrecht: Kluwer) 
 
\bibitem[Woosley (1993)]{Woosley93} 
  Woosley, S.E. 1993, \apj, 405, 273 

\bibitem [Zdziarski \& Lightman (1985)]{ZL85}
 Zdziarski, A.A., \& Lightman, A.P. 1985, \apj, 294, L79

\bibitem[Zhang, Kobayashi \& M\'esz\'aros (2003)]{ZKM03} 
  Zhang, B., Kobayashi, S. \& M\'esz\'aros, P. 2003, \apj, 595, 950  
 
\bibitem[Zhang \& M\'esz\'aros (2001)]{ZM01} 
  Zhang, B. \& M\'esz\'aros, P. 2001, \apj, 559, 110 
   
\end{thebibliography}
\end{document}